\documentclass[fleqn,usenatbib]{mnras}

\usepackage{newtxtext,newtxmath}

\usepackage[T1]{fontenc}

\DeclareRobustCommand{\VAN}[3]{#2}
\let\VANthebibliography\thebibliography
\def\thebibliography{\DeclareRobustCommand{\VAN}[3]{##3}\VANthebibliography}

\usepackage{graphicx}	
\usepackage{amsmath}	

\usepackage{rotating}
\usepackage{caption}
\usepackage[subrefformat=parens]{subcaption}
\usepackage{natbib}
\usepackage{multirow}
\usepackage{threeparttable}
\usepackage[normalem]{ulem}
\usepackage{float}
\restylefloat{figure}
\usepackage{soul}
\usepackage{mwe}

\title[X-ray Polarimetry for Black Hole Spin Measurements]{X-ray Polarimetry as a Tool to Measure the Black Hole Spin in Microquasars: Simulations of IXPE Capabilities}

\author[R. Mikusincova et al.]{%
Romana Mikusincova,$^{1}$\thanks{E-mail: romana.mikusincova@uniroma3.it}
Michal Dovciak,$^{2}$
Michal Bursa,$^{2}$
Niccolo Di Lalla,$^{3}$
Giorgio Matt,$^{1}$
\newauthor
Jiri Svoboda,$^{2}$
Roberto Taverna,$^{4}$
Wenda Zhang$^{5}$
\\
$^{1}$Dipartimento di Matematica e Fisica, Universit\`{a} Roma Tre, via della Vasca Navale 84, I-00146 Roma, Italy\\
$^{2}$Astronomical Institute, Academy of Sciences of the Czech Republic, Bo\v{c}n\'{i} II 1401, CZ-14100 Prague, Czech Republic\\
$^{3}$Department of Physics and Kavli Institute for Particle Astrophysics and Cosmology, Stanford University, Stanford, California 94305,USA\\
$^{4}$Dipartimento di Fisica e Astronomia, Universit\`{a} degli Studi di Padova, Via Marzolo 8, 35131 Padova, Italy\\
$^{5}$National Astronomical Observatories, Chinese Academy of Sciences, 20A Datun Road, Beijing 100101, China
}

\date{Accepted XXX. Received YYY; in original form ZZZ}

\pubyear{2022}

\begin{document}
\label{firstpage}
\pagerange{\pageref{firstpage}--\pageref{lastpage}}
\maketitle

\begin{abstract}

Measurements of the angular momentum (spin) of astrophysical black holes are extremely important, as they provide information on the black hole formation and evolution. We present simulated observations of a X-ray binary system with the Imaging X-ray Polarimetry Explorer (IXPE), with the aim to study the robustness of black hole spin and geometry measurements using X-ray polarimetry. As a representative example, we used the parameters of GRS 1915+105 in its former unobscured, soft state. In order to simulate the polarization properties, we modeled the source emission with a multicolor blackbody accounting for thermal radiation from the accretion disk, including returning radiation. Our analysis shows that the polarimetric observations in the X-ray waveband will be able to estimate both spin and inclination of the system with a good precision (without returning radiation we obtained for the lowest spin $\Delta a \leq 0.4$ (0.4/0.998 $\sim$ 40 \%) for spin and $\Delta i \leq 30^\circ$ (30$^\circ$/70$^\circ$ $\sim$ 43\%) for inclination, while for the higher spin values we obtained $\Delta a \leq 0.12$ ($\sim$ 12 \%) for spin and $\Delta i \leq 20^\circ$ ($ \sim $ 29\%) for inclination, within 1$ \sigma $ errors). When focusing on the case of returning radiation and treating inclination as a known parameter, we were able to successfully reconstruct spin and disk albedo in $\Delta a \leq 0.15$ ($\sim$ 15 \%) interval and $\Delta$ albedo $\leq 0.45$ (45 \%) intervals within 1$ \sigma $ errors. We conclude that X-ray polarimetry will be a useful tool to constrain black hole spins, in addition to timing and spectral-fitting methods.
\end{abstract}




\section{Introduction}

X-ray observations are crucial in the study of accreting systems around a compact central object. They provide us with information on physical processes in the inner regions of such objects and are used to constrain the spin of the black holes (BHs) in Active Galactic Nuclei (AGN) and X-ray binaries \citep{Reynolds2021}.

BH X-ray binary systems are very variable and often transient sources, going through different spectral states. Regarding their time variability, luminosity and spectral properties, different accretion states are observed \citep{Remillard2006}. Two basic states are distinguished, namely the low/hard and high/soft states. The low/hard state is characterized by a hard X-ray coronal emission with the accretion disk usually truncated at a large radius. In the high/soft case, the disk radius extend down to the Innermost Stable Circular Orbit (ISCO), and the thermal disk component dominates the spectrum in the classical $2$--$10$ keV band. There are several methods used for determining the BH spin; in the case of BH binaries, modeling the thermal emission is one of the most common \citep{Narayan2008}. Assuming that the inner radius of the disk coincides with the ISCO as it occurs in the soft state (see e.g. \citealt{Reynolds2008}, \citealt{Shafee2008}, \citealt{Steiner2010}), spectral measurements can in fact be used to probe the BH spin. In fact, the radius of the ISCO depends on the BH spin, going from 6 gravitational radii ($R_{\rm g} = GM/c^2$, where $ c $ is the speed of light) for a static black hole to 1 $R_{\rm g}$ for a maximally rotating black hole (see e.g. \citealt{NT1973}).

The temperature on the surface of the accretion disk depends on the radius according to the following rule of proportion (\citealt{Shakura1973}):
\begin{equation}
\label{EquationBB1}
T_{\textrm{eff}} \propto R^{-\frac{3}{4}}
\end{equation}
The emitted radiation at each disk radius is assumed to be a blackbody spectrum defined as (\citealt{Frank2002})
\begin{equation}
\label{EquationBB2}
I_{\nu} = \frac{2h}{c^2} \frac{\nu^3}{e^{h\nu / k_B T_{eff}}-1} = B_{\nu}(T_{\textrm{eff}})
\end{equation}
where $ h $ is the Planck constant and $ k_{B} $ is the Boltzmann constant. $ B_{\nu} $ is the spectral radiance density at frequency $ \nu $. This description is, however, only satisfied for a razor-thin accretion disk. In fact, an ionized layer of gas is formed above and below the accretion disk. This feature manifests itself into the observed spectrum and we also detect non-blackbody contributions. For this reason, we have to apply the color correction factor to account for the deviations from the blackbody spectrum (\citealt{Shimura1995}, \citealt{Merloni2000}, \citealt{Davis2005})
\begin{equation}
\label{EquationBB3}
f_{\textrm{col}} = \dfrac{T_{\textrm{col}}}{T_{\textrm{eff}}}
\end{equation}
For the color-corrected blackbody spectrum, we then obtain:
\begin{equation}
\label{EquationBB4}
I_{\nu} = f^{-4}_{\textrm{col}} \: B_{\nu}(f_{\textrm{col}} \: T_{\textrm{eff}})
\end{equation}
Determining the ISCO from the thermal spectrum allows one to estimate the BH spin (\citealt{Zhang1997}, \citealt{Gierlinski2001}). However, in order to get a reliable BH spin constraint, we must know precisely the BH mass, inclination of the accretion disk and the distance of the system from the observer (\citealt{Zhang1997}).

X-ray polarimetry provides an alternative method to measure the black hole spin of X-ray binaries in soft state. In fact, strong gravity effects modify the polarization properties of radiation emitted by the disk, with, in particular, a rotation of the polarization plane. The effect is larger at small radii, where the emitted radiation is also harder. As a consequence, a variation of the polarization angle with energy is expected \citep{Connors1980, Dovciak2008,Li2009}.

On January 2017, the Imaging X-ray Polarimetry Explorer \citep[IXPE,][]{Weisskopf2022} was selected in the framework of the NASA Small Explorers program. This mission, a collaboration between NASA and the Italian Space Agency (ASI), was successfully launched on December 9, 2021. Furthermore, the Chinese Academy of Sciences plans to launch the enhanced X-ray Timing and Polarimetry (eXTP) mission \citep{Zhang2016} in 2027. These instruments will observe X-ray polarization in the 2-8 keV energy band, very well suited for observing thermal radiation in accreting black holes in X-ray binary systems. In fact, measuring black hole spin using the thermal radiation method via X-ray polarimetry is one of the core scientific goals of both missions.

In this paper, we investigate the polarization properties of an X-ray binary system, GRS 1915+105, in the soft state and the robustness of measuring BH spin, inclination and orientation of the source on the plane of the sky using simulated X-ray polarimetric measurements with IXPE. Even if the source has been in an obscured state after a transition in 2018 \citep[][and references therein]{Ratheesh2021}, we use its former, well studied soft state as representative of an X-ray binary system (possibly a new transient) in such state. In Section \ref{sec:simulation} we discuss the modeling of multicolor blackbody and introduce the numerical codes we used to reproduce the polarimetric properties of the observed radiation. In particular we simulate both the simple case in which all photons arrive directly to the observer (using the code \textsc{kynbb}) and a more complex one in which photons can return to the disk due to strong gravity effects and then reach the observer after reflection (using the code \textsc{kynbbrr}). In Section \ref{sec:results} we analyze the simulated data with the aim to reconstruct spin and geometry of the source. Discussion and Conclusions follow in Section \ref{sec:discussion}.

\section{Polarization features from black hole accretion disk} \label{sec:simulation}

In this section we first summarize the methodology used to perform our study and then review the codes used (\textsc{kynbb} and \textsc{kynbbrr}) in order to have a detailed understanding of the simulations presented later in the paper.

\subsection{\textsc{kynbb} code} \label{subsec:kynbb}

For the sake of simplicity, we start considering only the contribution of radiation which arrives directly to the observer without interacting with the disk after emission (i.e. direct radiation). To this aim we used the code \textsc{kynbb}\footnote{\label{note1}\url{https://projects.asu.cas.cz/stronggravity/kyn/tree/master\#kynbb}} code \citep{Dovciak2008}, which is part of the relativistic model package \textsc{kyn} \citep{Dovciak2004, Dovciak2004T}. The theoretical framework is based on a few assumptions, namely the Kerr metric to describe the space-time around the central BH and an optically thick, geometrically thin Keplerian accretion disk, characterized by the Novikov-Thorne \citep{NT1973} surface temperature profile. The disk surface is then assumed to be covered by a geometrically-thin, optically-thick atmosphere, in which the main source of opacity is electron scattering (see e.g. \citealt{Chandrasekhar1960}, \citealt{Dovciak2004T}, \citealt{Dovciak2008}).

For the purpose of this work, we used the following model parameters: spin $ a $, inclination $ i $, BH mass $ M_{\textrm{BH}} $, accretion rate $ \dot{M} $, Thomson optical depth of the disk atmosphere $ \tau $, orientation of the system on the plane of the sky $ \chi $ (i.e. orientation of the projected axis of the system, $ -90 < \chi < 90 $) and normalization factor $ N $ defined as $1/D^2_{10}$ (where $D_{10}$ is the distance in units of $10$ kpc). The model allows for the possibility to use some additional parameters (e.g. to define an obscuring structure). However, we did not operate with those in our analysis. The full list of the \textsc{kynbb} model parameters can be found in Table \ref{tab:model_pars}.

The local polarization properties of the accretion disc emission were computed assuming different optical depths of the scattering atmosphere. In particular, the infinite optical depth approximation by \cite{Chandrasekhar1960} is applied when values of $\tau \gtrsim 10$ are considered\footnote{It has indeed been found (see \citealt{Dovciak2008}) that the result for an infinite slab is already reached for $\tau=10$.} and pre-calculated tables, obtained with the Monte Carlo code {\sc stokes} (\citealt{Goosmann2007, Marin2012, Goosmann2014}), are used for smaller $\tau$.

Thomson scattering is used to compute the polarization properties, while the variation as a function of the photon energy is accounted for using the color correction factor $f_{\rm col}$. Due to symmetry reasons, the emerging polarization is either parallel or perpendicular to the sky projection of the disk symmetry axis. We define the polarization angle to be zero for polarization parallel to the system axis (i.e. $U=0$ and $Q>0$ in terms of Stokes parameters), and $90^\circ$ for the orthogonal case (i.e. $U=0$ and $Q<0$). The Stokes parameter $V$, which denotes circular polarization, is zero in all cases discussed in this paper, since Thomson scattering does not generate circular polarization (and the current X-ray polarimeters cannot even measure it).

\begin{table}
\centering
\caption{Model parameters of \textsc{kynbb}} 
\label{tab:model_pars}
\begin{tabular}{p{0.14\textwidth}p{0.14\textwidth}p{0.14\textwidth}}
\hline
Model Parameter & Value & Free/Frozen \\
\hline
Black Hole Spin & (see Table~\ref{tab:results}) & Free (see Section~\ref{sec:results_kynbb}) \\
Inclination Angle & 70 & Free (see Section~\ref{sec:results_kynbb}) \\
$ r_{\mathrm{in}} $ & 1 & Frozen \\
Switch for $ r_{\mathrm{in}} $ & 1 & Frozen \\
$ r_{\mathrm{out}} $ & 1000 & Frozen \\
$ \phi$ & 0 & Frozen \\
$ d\phi$ & 360 & Frozen \\
$ M_{\mathrm{BH}} $ & 14 & Frozen \\
$ \dot{M} $ & (see Table~\ref{tab:results}) & Frozen \\
$ f_{\mathrm{col}} $ & 1.7 & Frozen \\
$ \alpha$ & -6 & Frozen \\
$ \beta$ & 0 & Frozen \\
$ r_{\mathrm{cloud}} $ & 0 & Frozen \\
Redshift & 0 & Frozen \\
ntable & 80 & Frozen\\
nrad & 150 & Frozen\\
Division & 1 & Frozen \\
$ n\phi $ & 180 & Frozen \\
Smooth & 0 & Frozen \\
Stokes & 1 & Frozen \\
$ \chi $ & 0 & Free \\
$\tau$ & 11 & Frozen \\
nthreads & 2 & Frozen\\
Normalization & 0.826 & Frozen\\
\end{tabular}
\end{table}

\subsection{\textsc{kynbbrr} code \label{sec:kynbbrr}}

A complete description of spectral and polarization properties of radiation coming from X-ray binaries in the soft state cannot ignore the contribution of returning radiation, i.e. photons which are bent by strong gravity effects and forced to return to the disk surface, where they can be reflected and eventually reach the observer. Modeling returning radiation features first requires to calculate the photon trajectories in the vicinity of the central BH, where general relativistic effects are more important. To this aim we used the ray-tracing based, \textsc{sim5} code \textsc{selfirr} (\citealt{Bursa2017}; \citealt{Zhang2019}). This code calculates all the null geodesics which end up on a given point on the disk surface from different incidence directions, starting from a different disk location. The disk surface is, then, divided into a number $\bar{N}_{\rm r}$ of incident points, each characterized by the distance $\bar{r}_{\rm i}$ from the central BH, while the possible incidence directions are sampled through a discrete $\bar{N}_\Theta\times\bar{N}_\Phi$ angular mesh, according to the polar angles $\bar{\Theta}_{\rm i}$ and $\bar{\Phi}_{\rm i}$ they form with the disk normal. For each geodesics, {\sc selfirr} returns the central distance $\bar{r}_{\rm e}$ and the emission direction angles $\bar{\Theta}_{\rm e},\bar{\Phi}_{\rm e}$, tracing back in this way all the possible returning photon trajectories. Alongside the numbers $\bar{N}_{\rm i}$, $\bar{N}_\Theta$ and $\bar{N}_\Phi$, which provide the grid dimensions, input parameters for each run are the BH mass $M$ and spin $a$. The output of the code is a fits file, containing all the geodesic parameters, which can be used inside the \textsc{kyn} package.

First simulations including returning radiation contributions assumed that returning photons were all reflected to the observer, i.e. with a $100$\% disk albedo \citep{Schnittman2009}. However, as shown in subsequent works \citep{Taverna2020}, in realistic conditions absorption may not be negligible and the albedo will be reduced correspondingly. As noted by \cite{Schnittman2009}, considering 100\% albedo would mean having a switch between PA values. At lower energies, direct radiation is more important, but at higher energies returning radiation becomes prevalent and has higher polarization degree. On the other hand, a constant albedo of 50\% would shift the switching energy to higher values. Following the work by \cite{Taverna2020}, where a self-consistent simulation for the albedo profile as a function of the energy was obtained using the software CLOUDY \citep{Ferland2017}, we fixed the albedo at the value of 50\%, which closely resembles that simulation (see their Figure 12). 

\begin{figure*}
        \centering
        \begin{subfigure}[b]{0.475\textwidth}
            \centering
            \includegraphics[width=\textwidth]{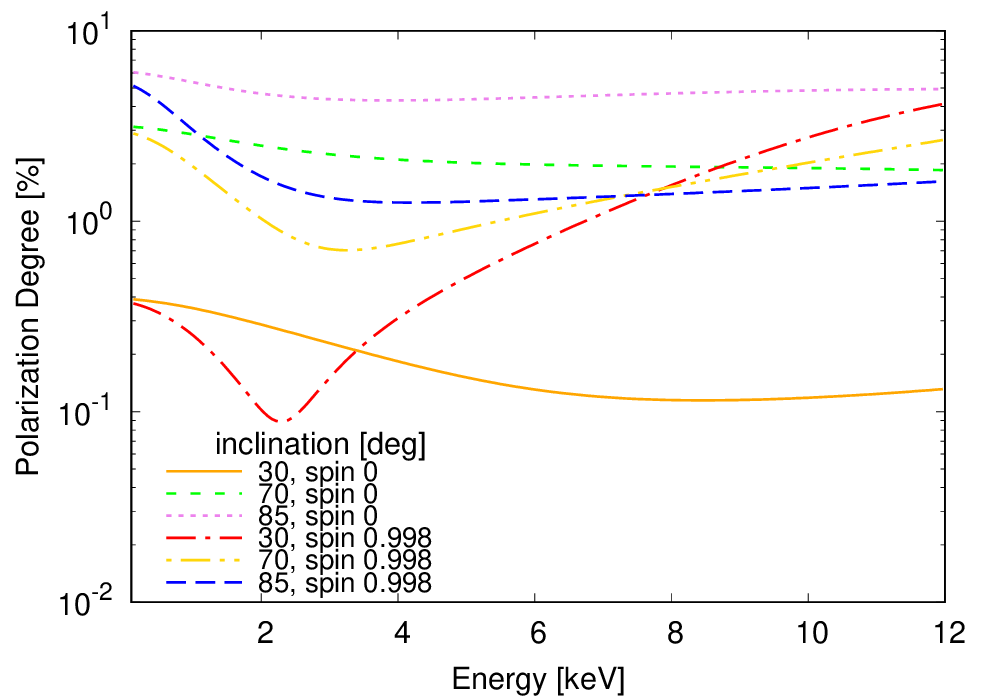}
            \caption[]%
            {{\small }}    
            \label{}
        \end{subfigure}
        \hfill
        \begin{subfigure}[b]{0.475\textwidth}  
            \centering 
            \includegraphics[width=\textwidth]{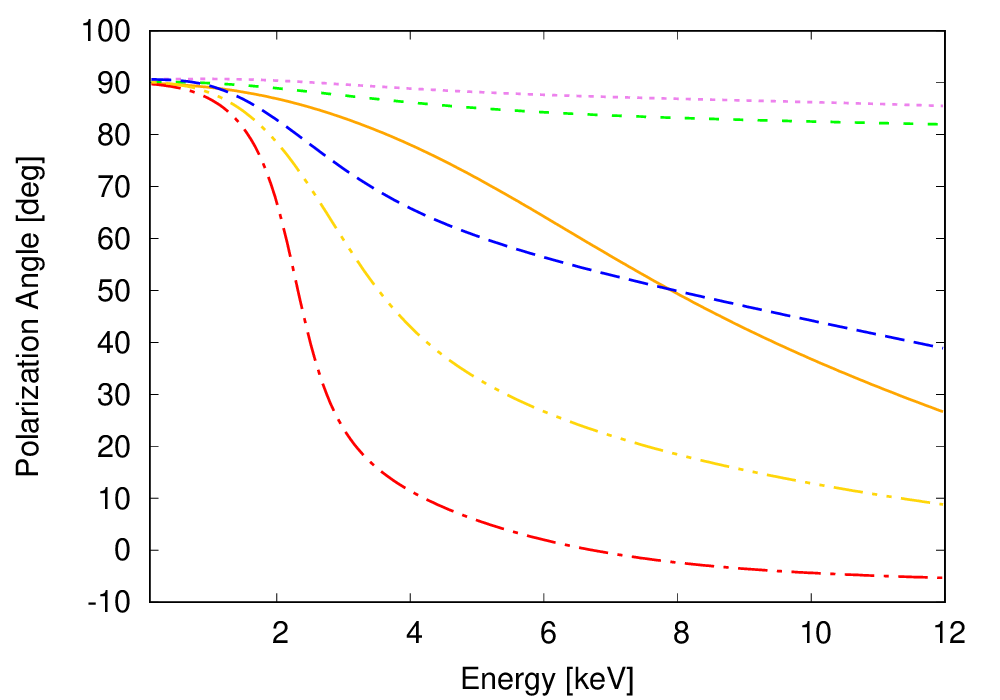}
            \caption[]%
            {{\small }}    
            \label{}
        \end{subfigure}
        \caption{Polarization degree (left) and polarization angle (right) dependence on energy for different inclination cases for GRS1915+105 with BH mass $ M = 14 $ M$ _{\odot} $ and spin a = 0 (orange, green and violet lines) and a = 0.998 (red, yellow and blue lines). The accretion rate is considered in such a way, that it would correspond to the observed luminosity $ L = 0.45 $ L$ _{\textrm{Edd}} $. The plots were created using the \textsc{kynbbrr} model.} 
        \label{fig:KYNBBRR_arate1}
\end{figure*}

\begin{figure*}
        \centering
        \begin{subfigure}[b]{0.475\textwidth}
            \centering
            \includegraphics[width=\textwidth]{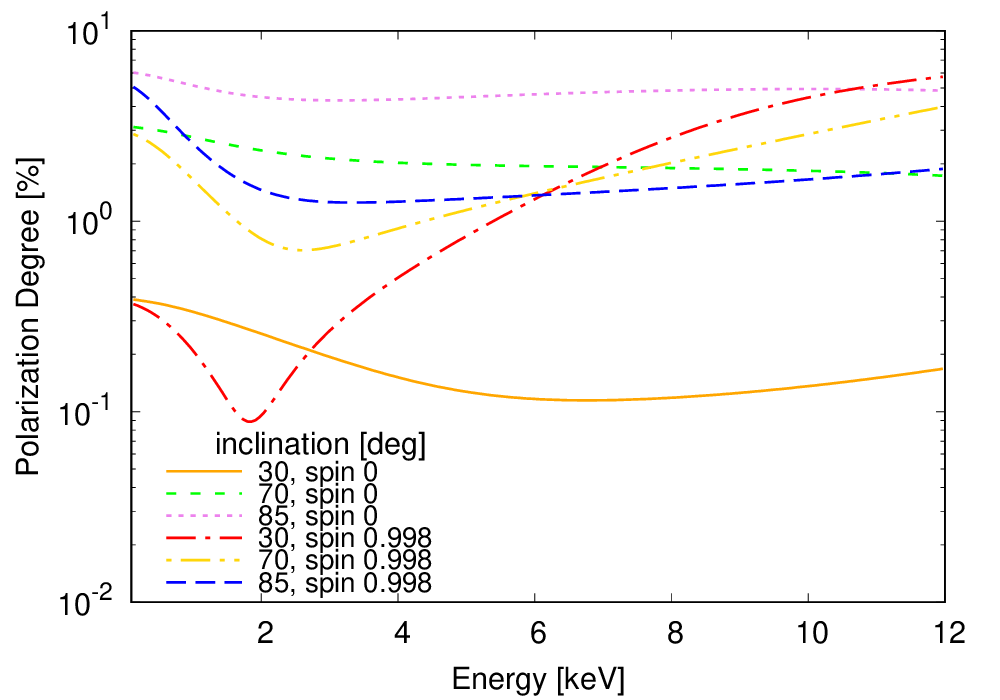}
            \caption[]%
            {{\small }}    
            \label{}
        \end{subfigure}
        \hfill
        \begin{subfigure}[b]{0.475\textwidth}  
            \centering 
            \includegraphics[width=\textwidth]{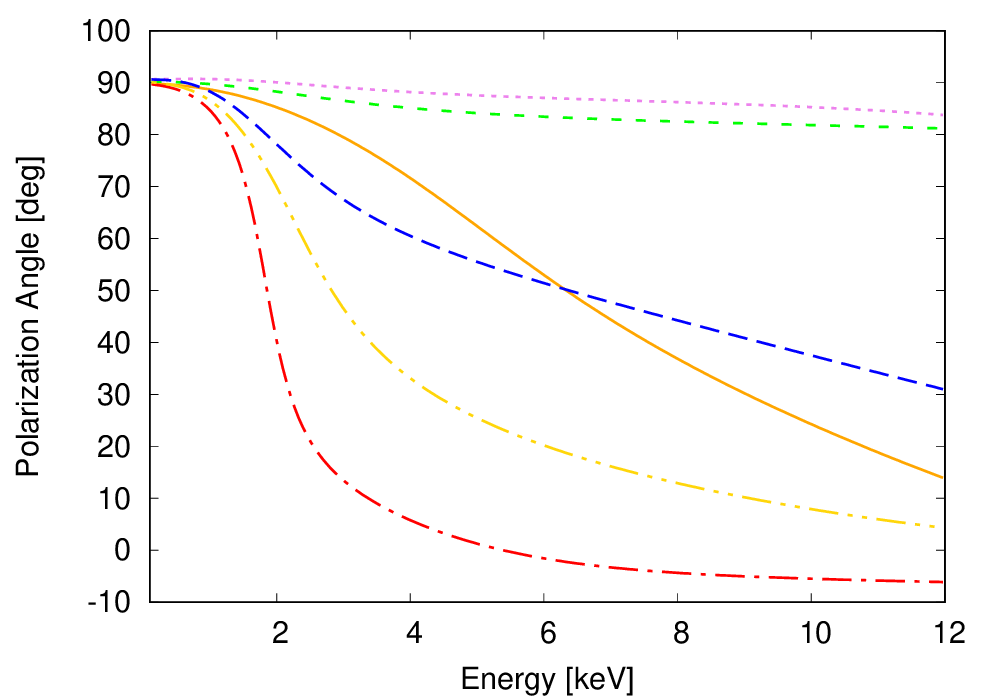}
            \caption[]%
            {{\small }}    
            \label{}
        \end{subfigure}
        \caption{Same as in Figure \ref{fig:KYNBBRR_arate1}, but with the accretion rate corresponding to the luminosity $ L = 0.185 $  L$ _{\textrm{Edd}} $. The plots were created using the \textsc{kynbbrr} model.}
        \label{fig:KYNBBRR_arate2}
\end{figure*}

While we observe a substantial change in PA across the $1$--$10$ keV band for all the inclinations for the highest spin values (bottom right panels of Figures \ref{fig:KYNBBRR_arate1} and \ref{fig:KYNBBRR_arate2}), this is only true for lower inclinations in the case of $a = 0$. Similarly, we observe a change in the behavior of PD, which tends to decrease at low energies, but then, starting at $\approx 2$ keV starts getting constant or slightly increasing. For the lowest inclination ($10^{\circ}$) we see a very pronounced minimum in PD, which is shifted towards lower energies for higher spins.

\subsubsection{Energy dependence of polarization features}

\begin{figure*}
        \centering
        \begin{subfigure}[b]{0.48\textwidth}
            \centering
            \includegraphics[width=\textwidth]{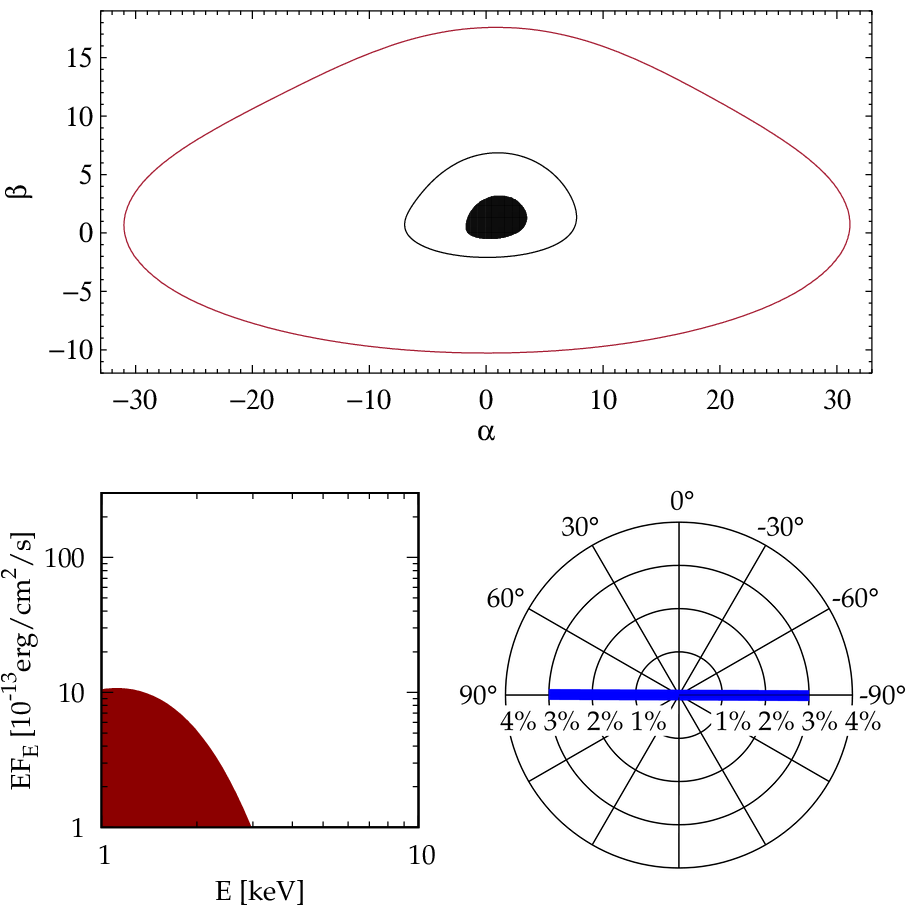}
            \caption[]%
            {{\small }}    
        \end{subfigure}
        \hfill
        \begin{subfigure}[b]{0.48\textwidth}  
            \centering 
            \includegraphics[width=\textwidth]{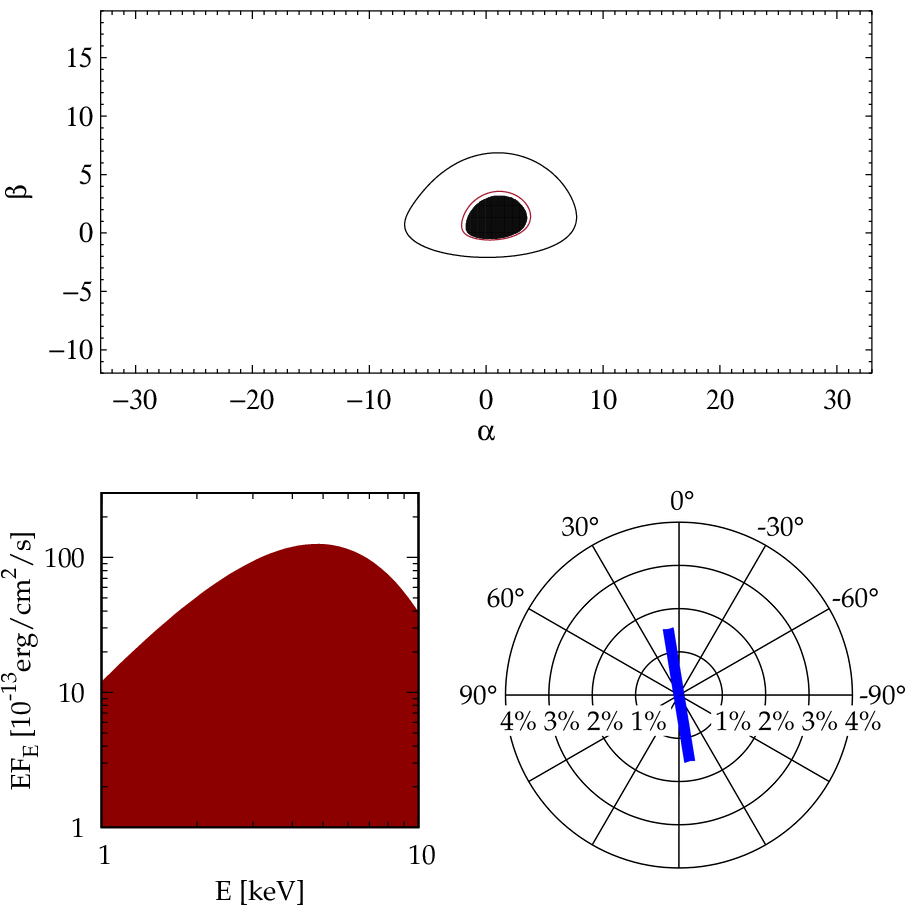}
            \caption[]%
            {{\small }}    
        \end{subfigure}
        \caption{Upper panel: representation of a black hole accretion disk on the sky of the observer. Bottom: flux and the polarization direction of collected radiation. The effects of variation of the polarization angle due to strong gravity are shown going from left to right. The figures represent a maximally rotating black hole. Starting from a region distant from the black hole, at $r=30r_{\rm g}$, (a), emission peaks at lower temperatures ($ \approx $ 1 keV) and the polarization is horizontal. Then, going very close to the black hole, at $r=1.285r_{\rm g}$, (b), the energy flux peaks at a higher energy ($ \approx $ 5 keV) and the polarization (blue line on the bottom right subpanel) is close to vertical. The length of the blue line represents the polarization degree, which gets lower closer to the source.} 
        \label{fig:PA_rotation}
\end{figure*}

The numerical code we used integrates the emission over the entire accretion disk, allowing us to shift from local to global polarization properties. At large distances from the black hole, the polarization degree is high (as represented by the length of the blue line on the bottom right subpanel of Figure \ref{fig:PA_rotation}), the polarization vector is parallel to the disk, and the emission is soft. Nearing close to the black hole, the emission is much harder and we observe a change in the direction of the polarization vector by about 30$ ^\circ $ (blue line on the bottom right subpanel). The polarization degree, represented by the length of the line, is smaller due to the superposition of different polarization vectors. These relativistic effects play an important role, as it can be observed in Figure \ref{fig:PA_rotation}.

\begin{figure*}
        \centering
        \begin{subfigure}[b]{0.475\textwidth}
            \centering
            \includegraphics[width=\textwidth]{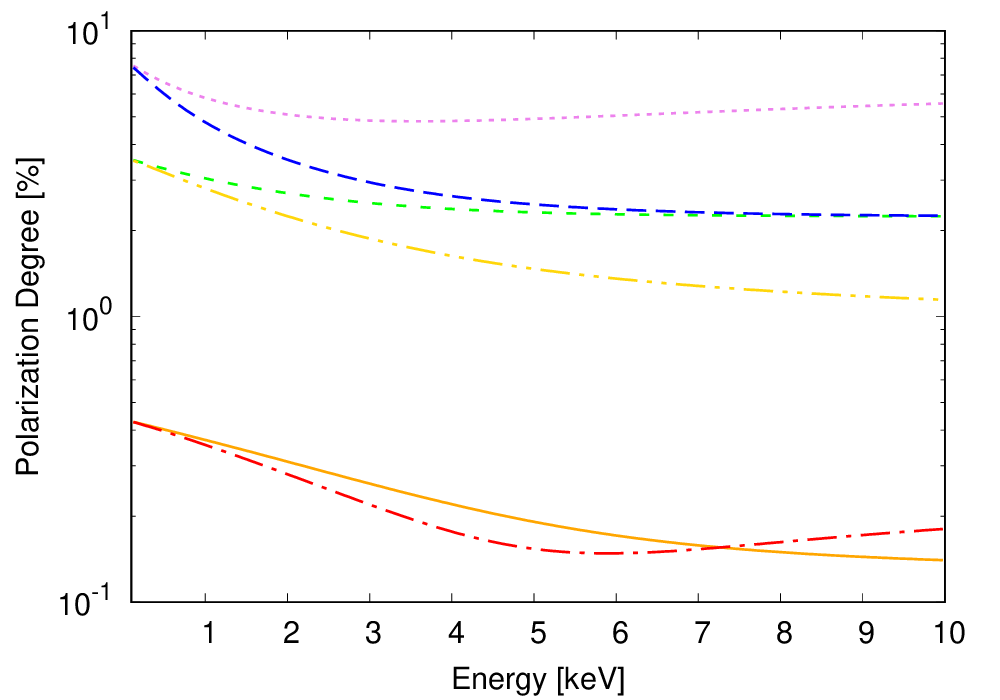}
            \caption[]%
            {{\small }}    
            \label{fig:}
        \end{subfigure}
        \hfill
        \begin{subfigure}[b]{0.475\textwidth}  
            \centering 
            \includegraphics[width=\textwidth]{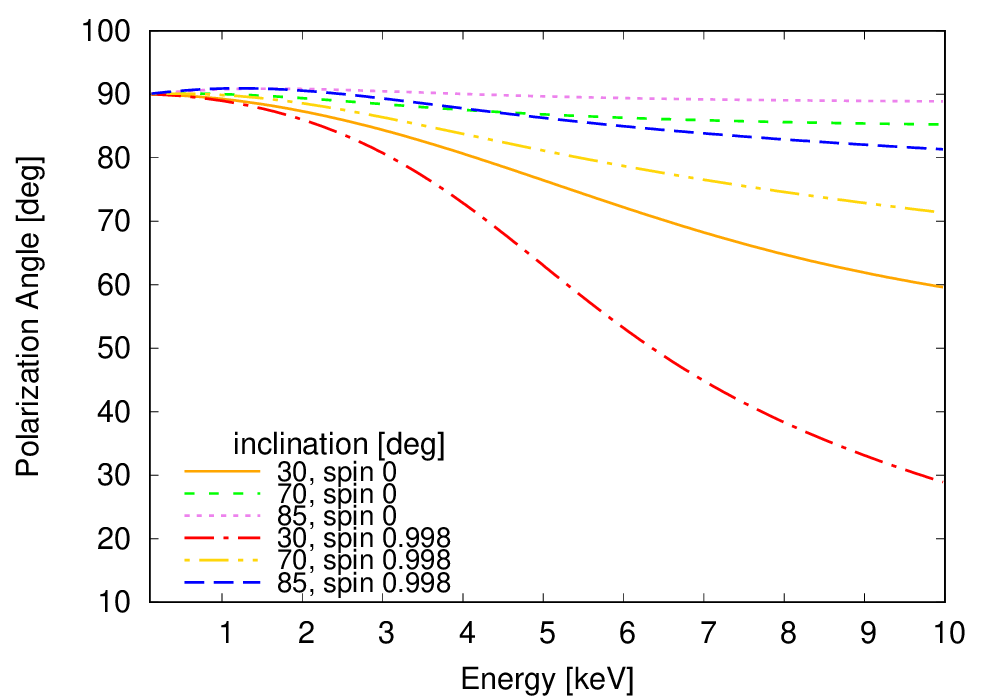}
            \caption[]%
            {{\small }}    
            \label{fig:}
        \end{subfigure}
        \caption{Same as in Figure \ref{fig:KYNBBRR_arate1}, but with the accretion rate corresponding to the luminosity $ L = 0.446 $ L$_{\textrm{Edd}} $. The plots were created using the \textsc{kynbb} model.}
        \label{fig:KYNBBincl_arate1}
\end{figure*}

\begin{figure*}
        \centering
        \begin{subfigure}[b]{0.475\textwidth}
            \centering
            \includegraphics[width=\textwidth]{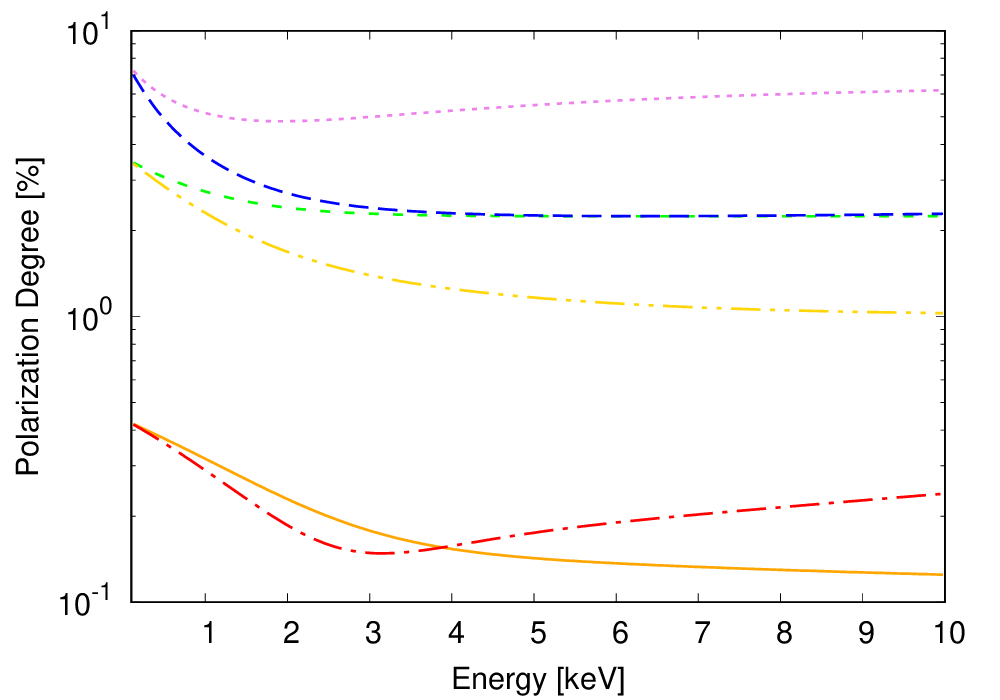}
            \caption[]%
            {{\small }}    
            \label{fig:}
        \end{subfigure}
        \hfill
        \begin{subfigure}[b]{0.475\textwidth}  
            \centering 
            \includegraphics[width=\textwidth]{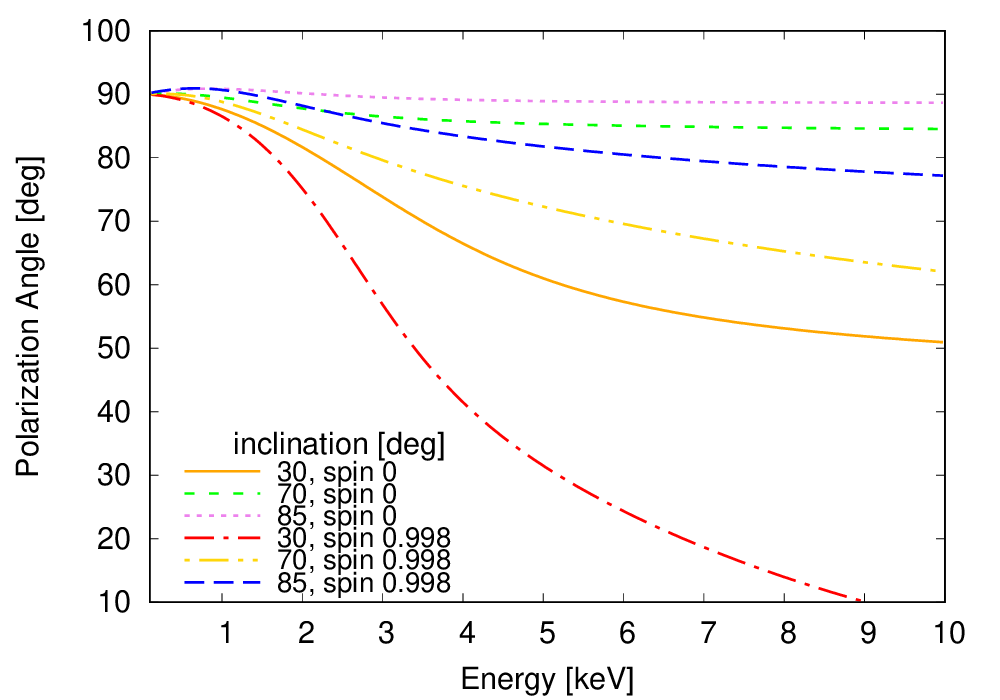}
            \caption[]%
            {{\small }}    
            \label{fig:}
        \end{subfigure}
        \caption{Same as in Figure \ref{fig:KYNBBRR_arate1}, but with the accretion rate corresponding to the luminosity $ L = 0.204 $ L$_{\textrm{Edd}} $. The plots were created using the \textsc{kynbb} model.}
        \label{fig:KYNBBincl_arate2}
\end{figure*}

Figures \ref{fig:KYNBBincl_arate1} and \ref{fig:KYNBBincl_arate2} (and similarly also Fig. \ref{fig:KYNBBRR_arate1} and Fig. \ref{fig:KYNBBRR_arate2}, but with the effects of returning radiation taken into account) show the dependence of the polarization degree (PD) and angle (PA) on energy for various inclination angles, two different values of the BH spin ($a = 0$ and $0.998$) and for different accretion rate values. These accretion rate values are set in such a way that the corresponding luminosity is $L = 0.446 L_{\textrm{Edd}}$ (Fig. \ref{fig:KYNBBincl_arate1}) and $L = 0.204  L_{\textrm{Edd}}$ (Fig. \ref{fig:KYNBBincl_arate2}) for the model without the returning radiation (\textsc{kynbb}) and luminosities $L = 0.45  L_{\textrm{Edd}}$ (Fig. \ref{fig:KYNBBRR_arate1}) and $L = 0.185 L_{\textrm{Edd}}$ (Fig. \ref{fig:KYNBBRR_arate2}) for the model with the effects of returning radiation into account (\textsc{kynbbrr}); $L_{\rm Edd}$ denotes the Eddington luminosity.

In the $2$--$8$ keV energy band the PD stays almost constant for both the spin and luminosity values sampled when returning radiation is not included, and for spin $a = 0$ with returning radiation. However, this is not the case for spin 0.998 when the impact of the returning radiation is taken into account (panels (c) in Figures \ref{fig:KYNBBRR_arate1} and \ref{fig:KYNBBRR_arate2}.). In these two cases, we see a rise of PD by $ \approx 1.5 \%$, accompanied by the steepest change in PA.

Generally, we observe very pronounced changes of PA in the cases of the highest spin value ($a = 0.998$) and for the modeled systems with lower inclination. This rapid change corresponds to the lowest PD, which is in agreement with the theoretical prediction (as described in the Figure \ref{fig:PA_rotation}). 
 
The polarization angle has a clear decreasing trend for $a = 0.998$, even for high inclination cases. On the other hand, it is nearly constant for higher inclination values over the $2$--$8$ keV band for spin $a = 0$ and both the accretion rate values in both models, while for low-inclination systems, it is distinctly decreasing. In the case of a non-rotating BH, the most pronounced change in PA is for $i = 10^{\circ}$ and $i = 30^{\circ}$ (all considered accretion rate values). The most prominent change is from $90^{\circ}$ down to $ \sim 15^{\circ}$ or $ \sim 0^{\circ}$ (Fig. \ref{fig:KYNBBRR_arate2}) and to $ \sim 55^{\circ}$ and $ \sim 20^{\circ}$ in the Fig. \ref{fig:KYNBBincl_arate2} for the source with and without returning radiation, respectively. Moreover, a slightly higher polarization is observed for a lower BH spin. This is caused by the rotation of the polarization angle in the strong gravitational field.

To summarize, for a non-spinning BH, where the inner disk radius is large, the polarization angle does not rotate much. For a spinning BH, where the inner disk radius is small, the polarization angle of the photons coming from the inner parts of the disk is much rotated, causing depolarization. Therefore, the net polarization is lower in the case of a higher BH spin.

\subsection{Data simulation}

\subsubsection{\textsc{IXPEobssim}}

\textsc{IXPEobssim} \citep{Baldini2022} is a Python-based software developed for X-ray polarimetric simulations of IXPE observations. The code provides the possibility of realistic simulations of observations based on the source model characteristics, and is able to simulate point-like, as well as extended sources.

As an input for the simulation code we used the data file generated by the \textsc{kynbb} (or \textsc{kynbbrr}) model containing information about energy, flux and Stokes parameters. Next, we used the configuration file within the code, which is built based on the source properties and morphology, and defined the energy band within which we performed the simulation ($2$--$8$ keV), the energy spectrum and the energy dependent polarization properties.

As a next step, we used a macrofile, which required a specification of the simulated observation exposure time and the energy binning. We simulated an IXPE observation with a $500$ ks exposure time. Since X-ray polarimetry requires a great number of photons to get a significant measurement, in order to have a sufficiently good statistic in each bin of the whole studied energy band ($2$--$8$ keV) we opted for the following binning in the plots of PD and PA: $2$--$2.5$, $2.5$--$3$, $3$--$3.5$, $3.5$--$4$, $4$--$4.5$, $4.5$--$5$, $5$--$6$, $6$--$8$ keV.

The spectra of the Stokes parameters ($ I $, $ Q $ and $ U $) are given in terms of \textsc{pha1}, \textsc{pha1q} and \textsc{pha1u} fits files, which are readable inside the software package \textsc{xspec}\footnote{https://heasarc.gsfc.nasa.gov/xanadu/xspec/}; then, the behaviors of polarization degree and angle as a function of energy are obtained reprocessing the Stokes parameter spectra through the polarization cube \textsc{pcube} algorithm.

\subsubsection{GRS 1915+105}

We performed a set of models of X-ray polarimetric data of GRS 1915+105, assuming different values for the black hole spin ($a = 0$, $0.7$, $0.9$ and $0.998$). Each data set was created in the $1$--$10$ keV energy band to encompass the energy band of the polarimeters, and consist of 161 linearly distributed data points for the $I$, $Q$ and $U$ Stokes parameters.

For the data sets, we also added the \textsc{TBabs} model accounting for the Galactic absorption with the column density $ n_{\textrm{H Galactic}}=1.39 \times 10^{22} $ cm$^{-2} $ (\citealt{Kalberla2005}). We used a flux value of $ F_{\textrm{2 - 8 keV}}=10^{-8} $ erg cm$^{-2}$ s$^{-1}$ (\citealt{Martocchia2006}), since this value was obtained from an observation of GRS 1915+105 in the soft state, which we intend to study. Therefore, we aimed to model this specific flux value for all of the cases studied, which is why we had to change the accretion rate $\dot{M}$ for each modeled spin value.

We assumed the distance of the source to be $D=11$ kpc, as follows from the observation by \cite{Jonker2004} \footnote{Newer results by \cite{Reid2014} suggest a distance of $ 8.6^{+2}_{-1.6} $ kpc, broadly compatible with our choice within 1 $ \sigma $.}. So, in our case we obtain $ N=0.826$.

Regarding the inclination angle, we left it as a varying parameter in order to use our calculations for a general source in the soft state. However, in the case of GRS 1915+105 the inclination has been measured to be $66^{\circ}\pm2^{\circ}$ \citep{Fender1999} and this information can be used to reduce the uncertainties on the other parameters. The same can be said of the orientation, as far as we can assume that the disk symmetry axis coincides with the direction of the observed jet.

As the final output of the data simulation procedure, we get a dataset containing flux values, Stokes parameters, polarization degree (PD) and polarization angle (PA), with their respective $1\sigma$ errors, as functions of the energy. Stokes $Q$ and $U$ were used as an input for simulations of a polarimetric observation with \textsc{IXPEobssim}. The output of the simulation is a set of data files which are ready to be used for an analysis within the \textsc{xspec} software.

\begin{figure*}
        \centering
        \begin{subfigure}[b]{0.475\textwidth}
            \centering
            \includegraphics[width=\textwidth]{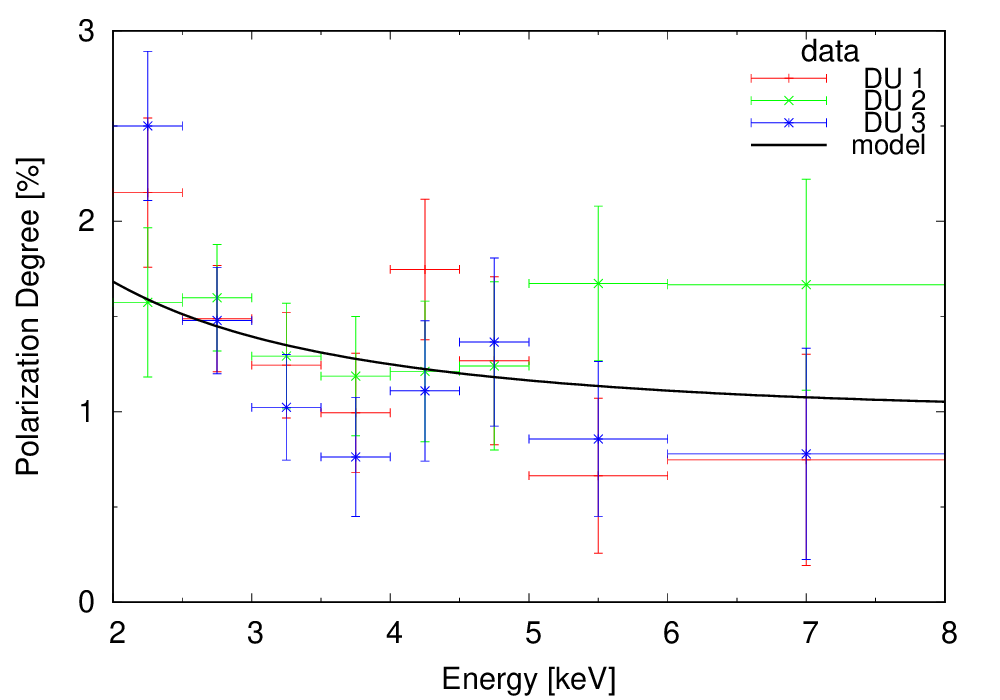}
            \caption[]%
            {{\small }}    
            \label{}
        \end{subfigure}
        \hfill
        \begin{subfigure}[b]{0.475\textwidth}  
            \centering 
            \includegraphics[width=\textwidth]{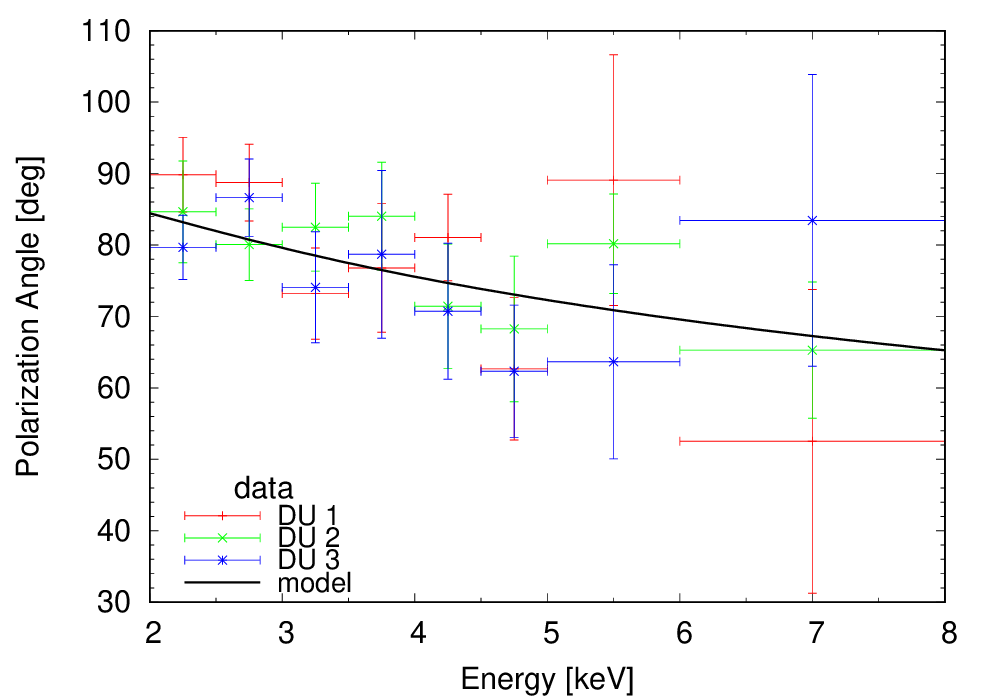}
            \caption[]%
            {{\small }}    
            \label{}
        \end{subfigure}
        \vskip\baselineskip
        \begin{subfigure}[b]{0.475\textwidth}   
            \centering 
            \includegraphics[width=\textwidth]{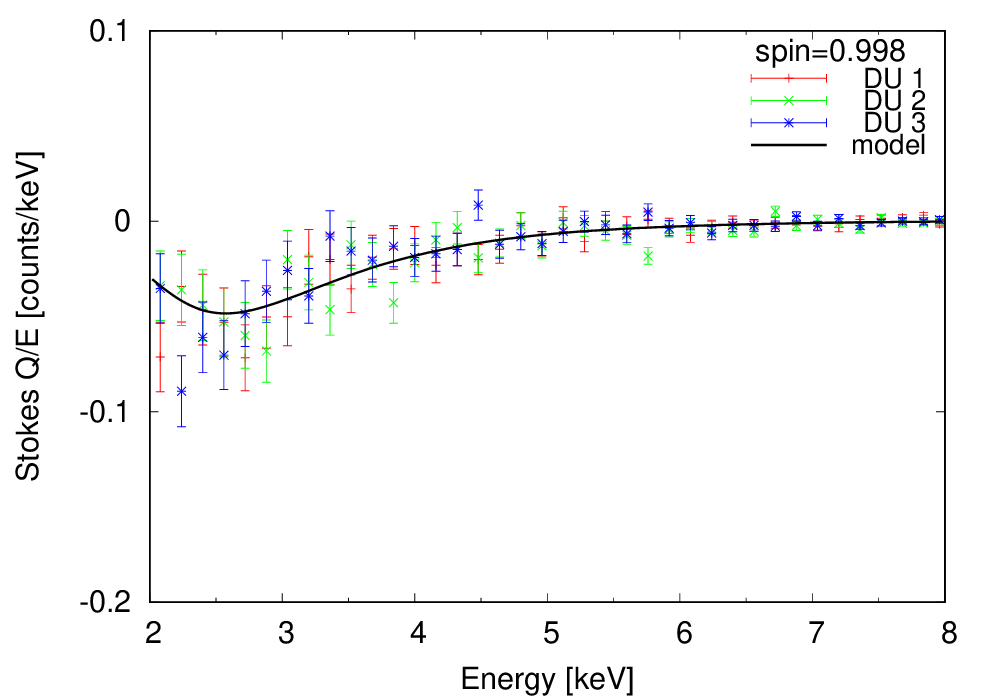}
            \caption[]%
            {{\small }}    
            \label{}
        \end{subfigure}
        \hfill
        \begin{subfigure}[b]{0.475\textwidth}   
            \centering 
            \includegraphics[width=\textwidth]{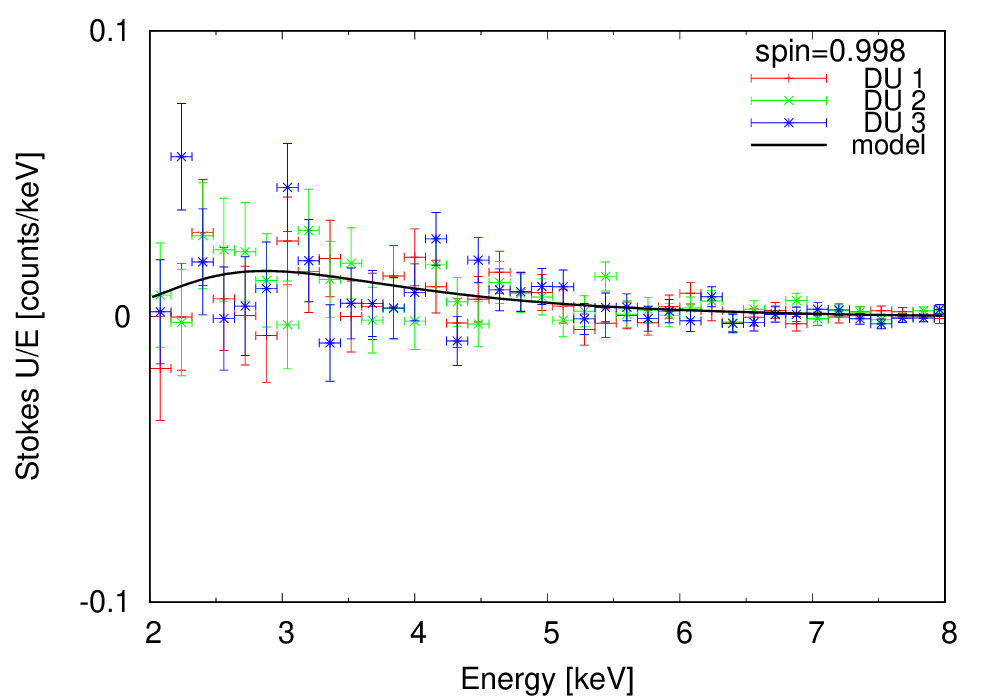}
            \caption[]%
            {{\small }}    
            \label{}
        \end{subfigure}
        \caption{Energy dependence of simulated data for Polarization Degree (a) and Angle (b); energy dependence of Stokes parameter Q/E (c) and U/E (d) for spin value $a=0.998$, inclination $i=70^\circ$, and optical depth $\tau \rightarrow \infty$. The solid black line represents theoretical model. The data were simulated using a model without returning radiation (\textsc{kynbb}).}
        \label{fig:simulated}
\end{figure*}

Figure \ref{fig:simulated} shows simulated data for energy dependence of PD (a), PA (b), specific Stokes parameters $ Q/E $ (c) and $ U/E $ (d). Each simulation contains three sets of data, one per each detector unit of IXPE. In (a) we present PD, which stays relatively constant through the 2-8 keV interval (except for the very low energy interval), while PA (in (b)) has a rather decreasing trend from $ \sim 85^\circ $ to $ \sim 65^\circ $. While the parameter $ U/E $ stays almost constant through the entire energy band, parameter $ Q/E $ presents a decreasing trend and is correlated with the similar trend of flux dependence. In particular, the behavior of $ Q/E $ and $ U/E $ have a substantial deviation from 0 only at lower energies (2--4 keV), negative for $ Q/E $ and positive for $ U/E $.

\section{Results} \label{sec:results}

To see whether we can reliably reconstruct model parameters from polarization signatures, we inspected the simulated polarization properties. We started exploring the effects of varying the BH spin and the inclination and orientation of the disk for direct radiation, using the \textsc{kynbb} code.

Then, we considered in addition returning radiation effects with simulations based on \textsc{kynbbrr}. In the latter case, to better highlight the effects of returning radiation, we assumed the inclination and orientation of the source to be known parameters.

We chose to work with the Stokes parameters instead of PD and PA. Stokes parameters are independent quantities, while PD and PA are not, being derived
from the Stokes parameters $I$, $Q$ and $U$ by the following relations (for linear polarization):
\begin{subequations}
\label{Equation5}
\begin{align}
\mathrm{PD} &= \dfrac{\sqrt{Q^{2}+U^{2}}}{I} \\
\mathrm{PA} &= \dfrac{1}{2} \arctan\left(\dfrac{U}{Q} \right) 
\end{align}  
\end{subequations}
which are non-linear and, therefore, PD and PA will bear non-gaussian errors.

\begin{table}
    \setlength{\tabcolsep}{3pt}
	\centering
	\caption{ Best-fit values for the spin ($a$), inclination ($i$) and orientation of the source on the sky of the observer ($ \chi $) for simulated exposure times $500$ ks, $250$ ks and $125$ ks, modeled and fitted with \textsc{kynbb}. (u) - errors that were unconstrained during the fitting procedure.}
	\label{tab:results}
	\begin{tabular}{ccccc} 
		\hline
		 & Model & 500 ks & 250ks & 125 ks\\
		\hline
        & & $ \dot{M} [M_{\odot}/yr] = 2.43 \times 10^{-7} $  &  & \\
        \hline
		$a$ & 0 & 0.017$ ^{+0.383}_{-0.017} $ & 0.0$ ^{+0.11}_\textrm{-u} $ & 0.0$ ^{+0.16}_\textrm{-u} $ \\ \\
        $ i [ ^{\circ} ] $ & 70 & 71$ ^{+6}_{-21} $ & 76$ ^{+4}_{-28} $ & 77$ ^{+5}_{-34} $\\ \\
		$ \chi [ ^{\circ} ] $ & 0 & 0$ ^{+3}_{-2} $ & 0$ ^{+1}_{-3} $ & -3$ \pm 3 $ \\ \\
		$ \chi^{2}/d.o.f. $ & -- & 946.51/897 & 919.7/897 & 873/897 \\
		\hline
        & & $ \dot{M} [M_{\odot}/yr] = 8.92 \times 10^{-8} $   &  & \\
        \hline
		$a$ & 0.7 & 0.7$ ^{+0.09}_{-0.05} $ & 0.82$ ^{+0.11}_{-0.16} $ & 0.7$ ^{+0.08}_{-0.06} $ \\ \\
		$ i [ ^{\circ} ] $ & 70 & 68$ ^{+7}_{-8} $ & 59$ ^{+20}_{-8} $ & 77$ ^{+5}_{-21} $ \\ \\
		$ \chi [ ^{\circ} ] $ & 0 & 0$ ^{+3}_{-2} $ & 6$ ^{+4}_{-6} $ & -5$ ^{+7}_{-4} $ \\ \\
		$ \chi^{2}/d.o.f. $ & -- & 841.21/897 & 905.04/897 & 873.03/897 \\
		\hline
        & & $ \dot{M} [M_{\odot}/yr] = 4.70 \times 10^{-8} $   &  & \\
        \hline
		$a$ & 0.9 & 0.91$ ^{+0.068}_{-0.064} $ & 0.988$ ^\textrm{+u}_{-0.09} $ & 0.858$ ^{+0.06}_{-0.05} $ \\ \\
		$ i [ ^{\circ} ] $ & 70 & 66$ ^{+12}_{-8} $ & 58$ ^{+11}_{-4} $ & 79$ ^{+7}_{-25} $ \\ \\
		$ \chi [ ^{\circ} ] $ & 0 & 4$ ^{+5}_{-3} $ & 9$ ^{+4}_{-7} $ & -1$ ^{+14}_{-5} $ \\ \\
		$ \chi^{2}/d.o.f. $ & -- & 900.7/897 & 847.9/897 & 980.8/897 \\
		\hline
        & & $ \dot{M} [M_{\odot}/yr] = 1.98 \times 10^{-8} $   &  & \\
        \hline
		$a$ & 0.998 & 0.998$ ^\textrm{+u}_{-0.021} $ & 1.0$ ^\textrm{+u}_{-0.020} $ & 0.988$ ^\textrm{+u}_{-0.020} $ \\ \\
		$ i [ ^{\circ} ] $ & 70 & 69$ ^{+8}_{-3} $ & 71$ \pm 10 $ & 79$ ^{+8}_{-12} $ \\ \\
		$ \chi [ ^{\circ} ] $ & 0 & 1$ ^{+4}_{-6} $ & 1$ ^{+4}_{-7} $ & -8$ ^{+9}_{-6} $ \\ \\
		$ \chi^{2}/d.o.f. $ & -- & 915.2/897 & 860.29/897 & 865.5/897 \\
	\end{tabular}
\end{table}

\subsection{\textsc{kynbb} \label{sec:results_kynbb}}

\noindent
We studied the capability in constraining the spin of the black hole and the inclination of the accretion disk from the simulated observations by fitting the Stokes spectra leaving free these two parameters, as well as the orientation of the system.

\begin{figure*}
        \centering
        \begin{subfigure}[b]{0.33\textwidth}
            \centering
            \includegraphics[width=\textwidth]{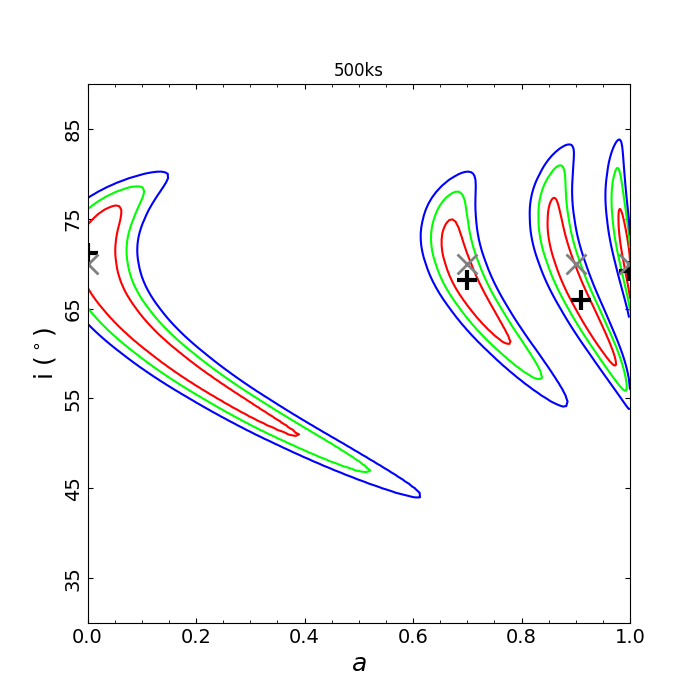}
            \caption[]%
            {{\small }}    
            \label{}
        \end{subfigure}
        \hfill
        \begin{subfigure}[b]{0.33\textwidth}  
            \centering 
            \includegraphics[width=\textwidth]{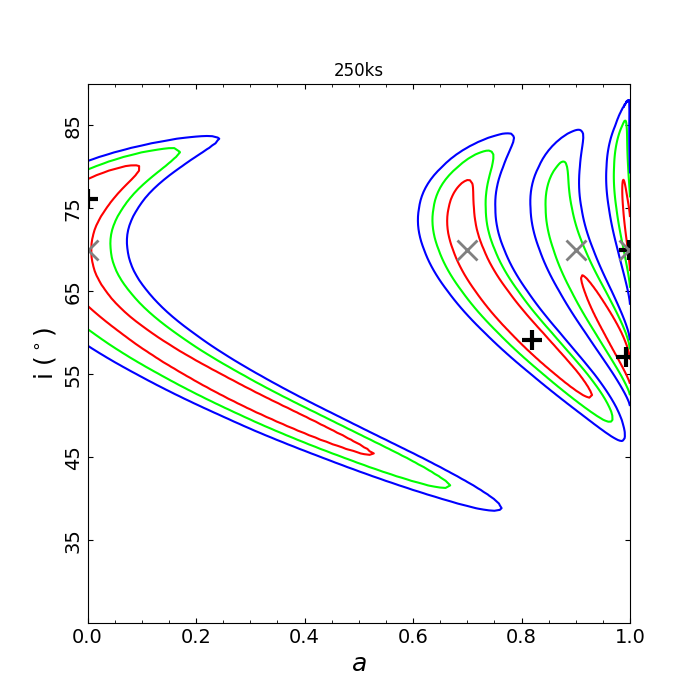}
            \caption[]%
            {{\small }}    
            \label{}
        \end{subfigure}
        \hfill
        \begin{subfigure}[b]{0.33\textwidth}  
            \centering 
            \includegraphics[width=\textwidth]{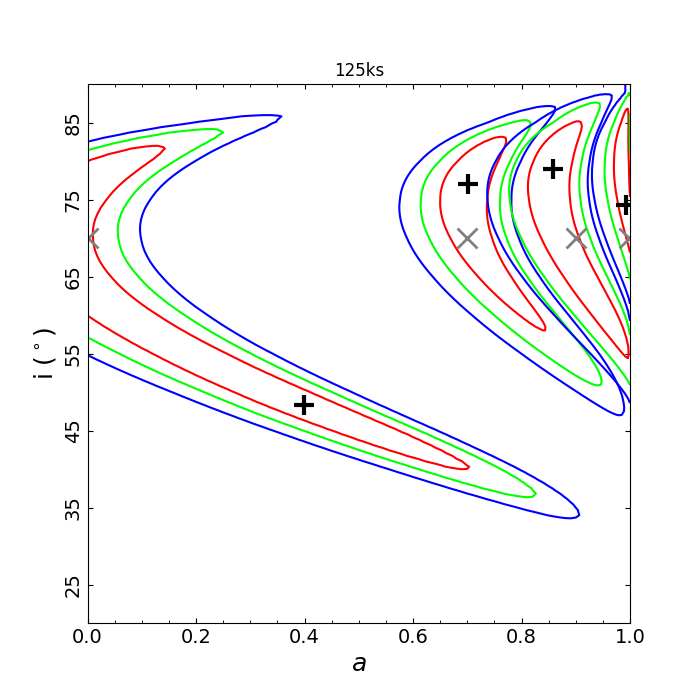}
            \caption[]%
            {{\small }}    
            \label{}
        \end{subfigure}
        \caption{Contour plots of spin vs inclination resulting from the fit the simulated Stokes parameters $Q$ and $U$ obtained for $a = 0$, $0.7$, $0.9$, $0.998$ for various exposure times: (a) $t = 500$ ks, (b) $t = 250$ ks and (c) $t = 125$ ks. Confidence contours at $1\sigma$ (red), $2\sigma$ (green) and $3\sigma$ (blue) are also shown. The gray cross marks the original values, which were (left to right, for each panel) spin a = 0, 0.7, 0.9 and 0.998 and i = 70 $ \circ $ for all the plotted cases. The black "plus" sign shows the best fit value, which is (again, left to right) a = 0, 0.7, 0.91 and 0.998 (a), a = 0, 0.82, 0.992 and 0.999 (b) and a = 0.4, 0.7, 0.86 and 0.993 (c). Data are plotted for a model without returning radiation (\textsc{kynbb}).}
        \label{fig:exposure}
\end{figure*}

The input and the best fit values for all of  studied cases are reported in Table \ref{tab:results}. We produced contour plots of spin versus inclination for all of the reported cases. We show contours for 1 $ \sigma $ (red), 2 $ \sigma $ (green) and 3 $ \sigma $ (blue) confidence levels in the Figure \ref{fig:exposure} panel (a), obtained for a 500 ks simulation.

\begin{figure*}
        \centering
        \begin{subfigure}[b]{0.48\textwidth}
            \centering
            \includegraphics[width=\textwidth]{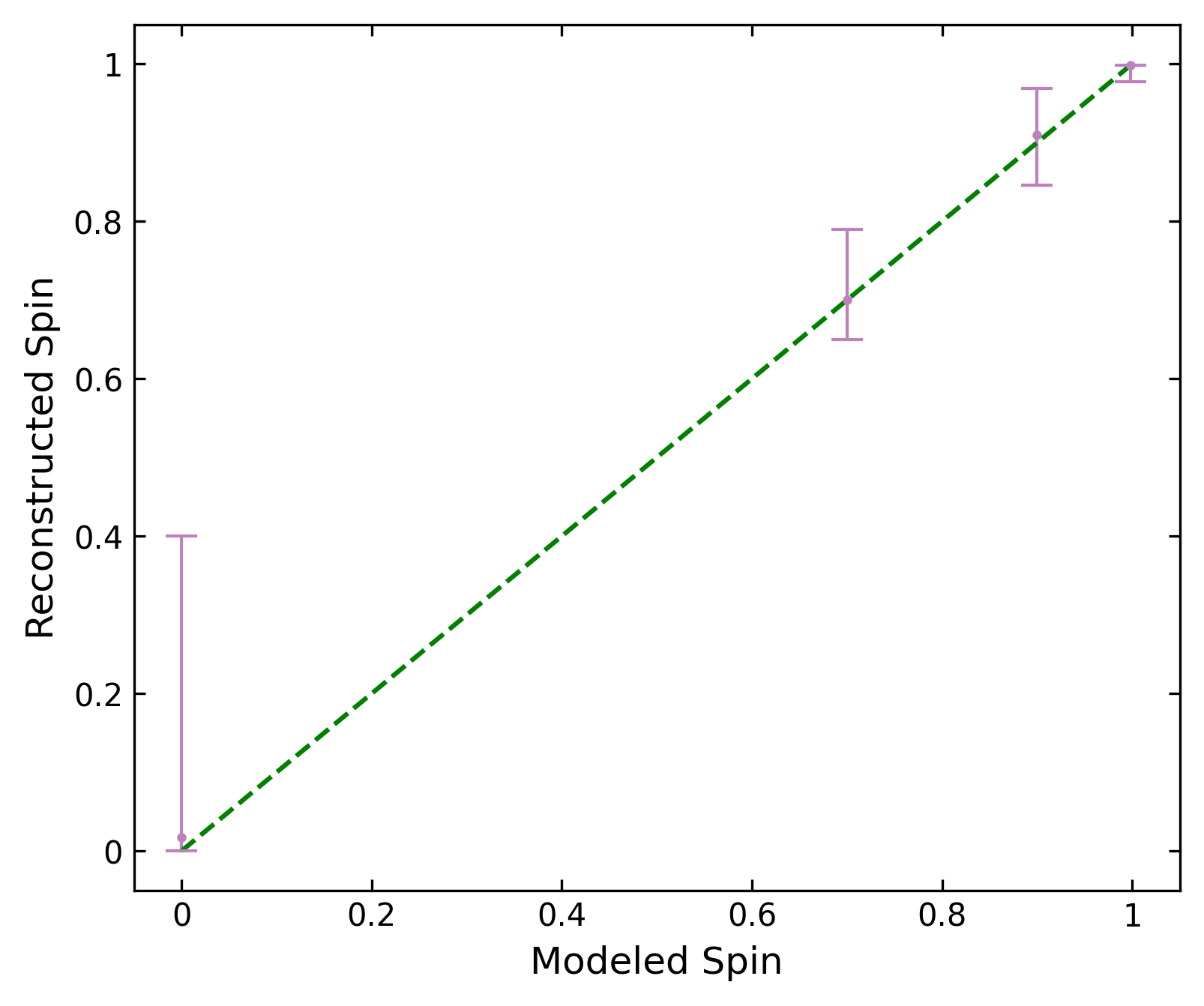}
            \caption[]%
            {{\small }}    
            \label{}
        \end{subfigure}
        \hfill
        \begin{subfigure}[b]{0.48\textwidth}  
            \centering 
            \includegraphics[width=\textwidth]{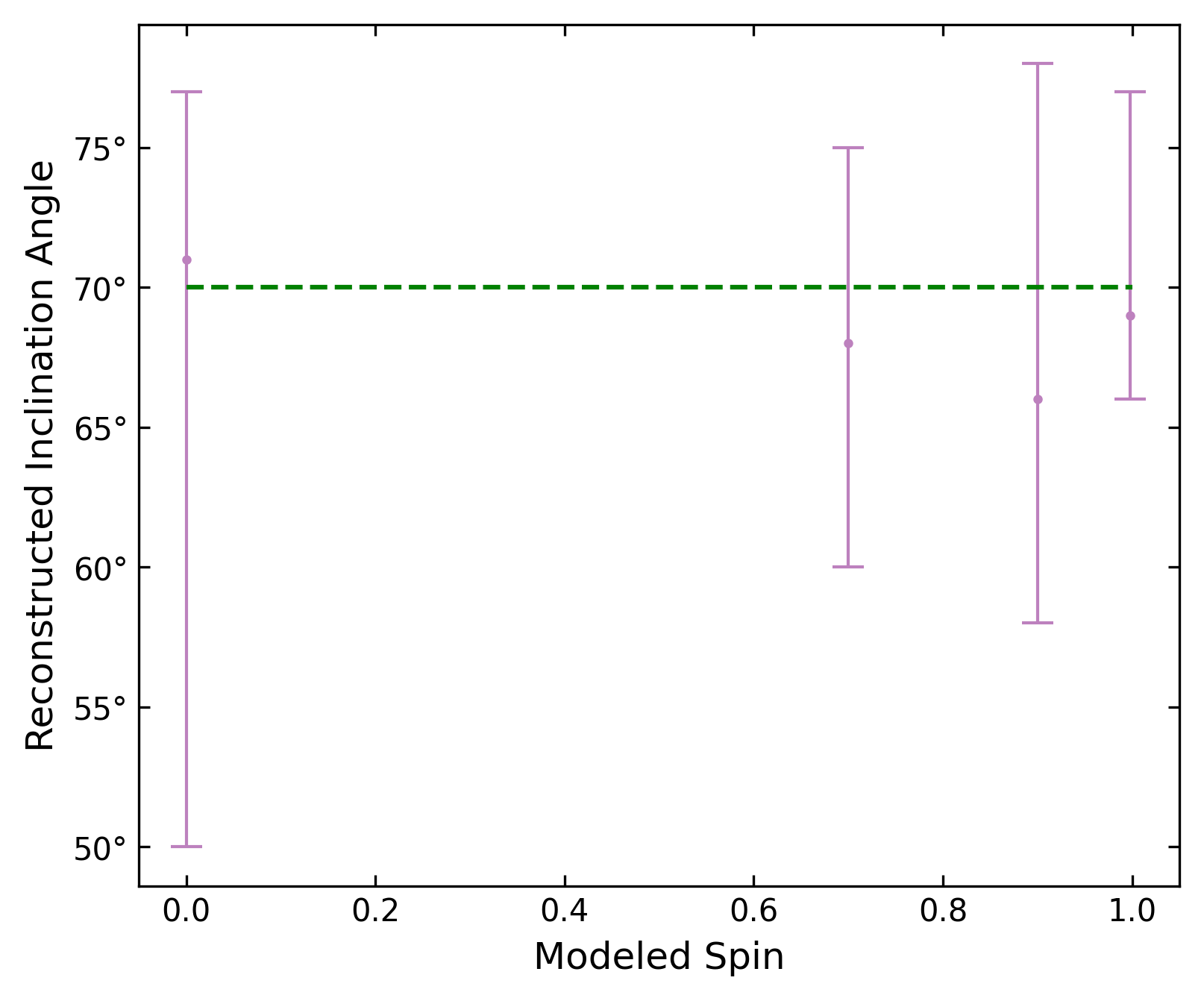}
            \caption[]%
            {{\small }}    
            \label{}
        \end{subfigure}
        \caption{Spin (a) and inclination (b) constraints with the \textsc{kynbb} model versus input values. Green dashed line represents the 1:1 relation for reconstructed and input spins (a) and the input value for the inclination (b).}
        \label{fig:goodnes}
\end{figure*}

The input parameters were correctly reconstructed,  as the gray cross (marking the original value) and black "plus" signs (best-fit) almost overlap for all four of the studied spin cases. The interval of good-fit values (considered here and later for 1$ \sigma $) for inclination is spread out through  $\Delta i \approx  20^\circ$ (20$^\circ$/70$^\circ$ $ \sim $ 29\%) for the cases of spin $a = 0.7$, $0.9$ and $0.998$, while the spin is constrained within $ \Delta a \lesssim  0.12$ ($\sim$ 12 \%). The case of $a = 0$ seems to be the most problematic to reconstruct, as we obtained $ \Delta i \sim  30^\circ$ ($\sim$ 43 \%) and  $ \Delta a \sim 0.4$ ($\sim$ 40 \%). We show the strength of the spin and inclination constraints in Figure \ref{fig:goodnes} (panels (a) and (b), respectively) as a function of the modeled values for these two parameters. The function is represented by the green dashed line. The reconstructed values lay on the green line for all the studied spin cases within their corresponding errors, which manifests the strength of the used method. We simulated the same case for different exposure times of $250$ ks and $125$ ks, shown in panels (b) and (c) of the Figure \ref{fig:exposure}. It is clear from the plots that, as expected, the lower the simulated exposure time, the higher the uncertainty in the spin and inclination reconstruction.

\subsection{\textsc{kynbbrr}}

In this part, we inspect the effects of returning radiation on the observed polarization properties of source radiation. We first had to use the \textsc{selfirr} code calculating the null geodesics through the accretion disk. We used a grid of dimensions $\bar{N}_{\rm i}$ = 500, $\bar{N}_\Theta$ = 100 and $\bar{N}_\Phi$ = 100, as described in Section \ref{sec:kynbbrr}. Then, we used the code \textsc{kynbbrr} in order to model the returning radiation effects of the source. To study this process, we treat the inclination and the orientation of the source as known parameters which, as discussed above, is a solid assumption for GRS 1915+105.  We simulated observations of GRS1915+105 for IXPE, as in the previous section and used the data of the Stokes parameters $Q$ and $U$ in $2$--$8$ keV interval.

\begin{figure*}
        \centering
        \begin{subfigure}[b]{0.48\textwidth}
            \centering
            \includegraphics[width=\textwidth]{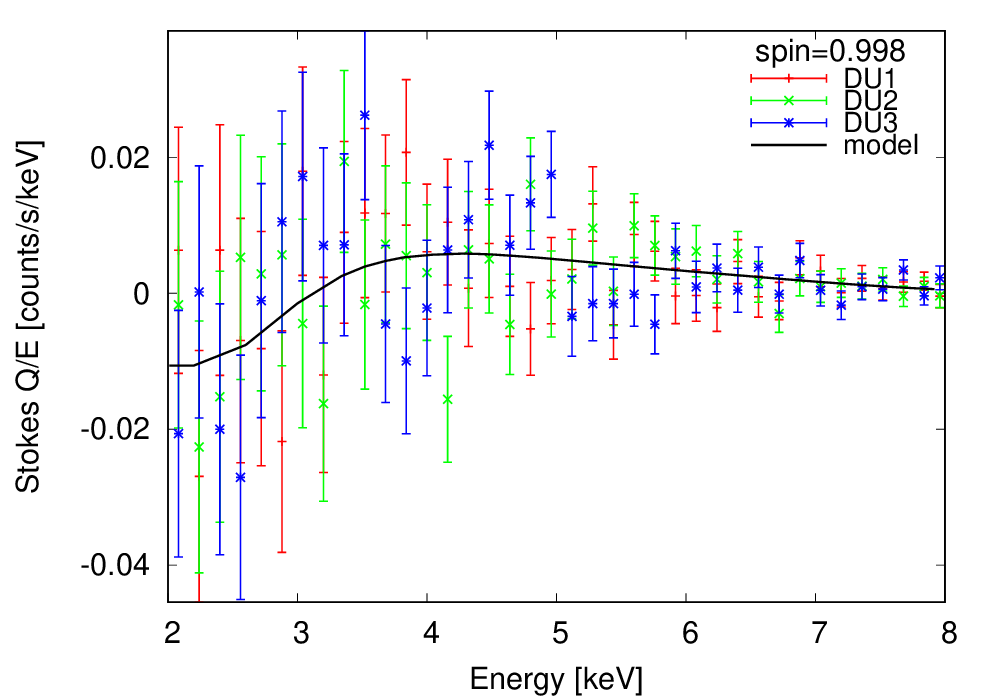}
            \caption[]%
            {{\small }}    
            \label{}
        \end{subfigure}
        \hfill
        \begin{subfigure}[b]{0.48\textwidth}  
            \centering 
            \includegraphics[width=\textwidth]{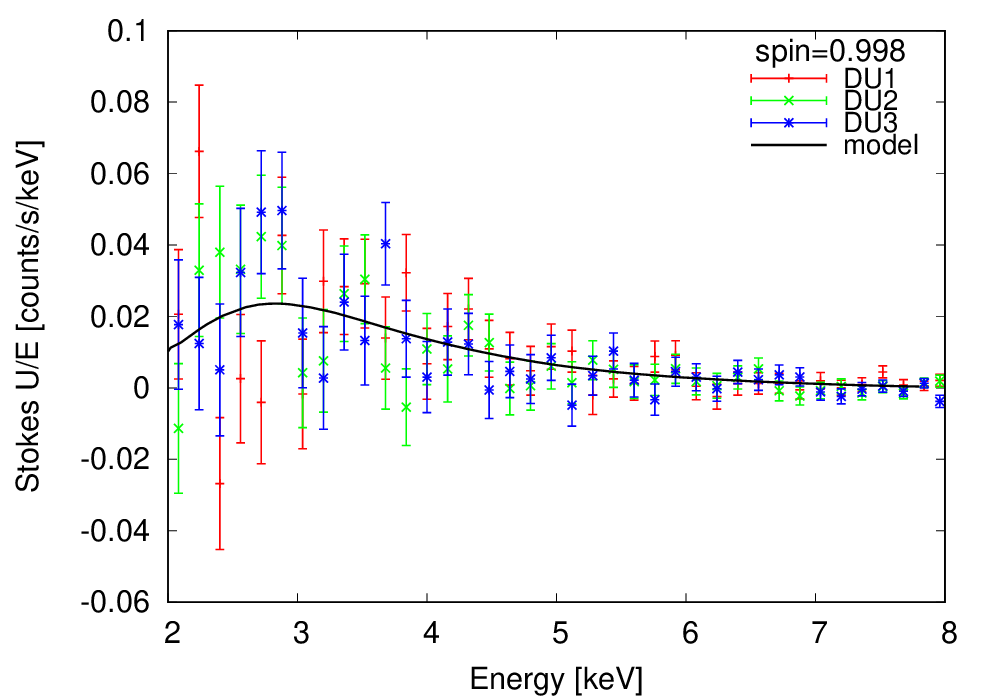}
            \caption[]%
            {{\small }}    
            \label{}
        \end{subfigure}
        \caption{Energy dependence of Stokes Q/E (a) and U/E (b) data simulated for spin $0.998$ and for each of the three IXPE detector units fitted with the theoretical model. Response matrix was applied on the models. The data were simulated using a model with returning radiation (\textsc{kynbbrr}).}
        \label{fig:kynbbrr_data}
\end{figure*}

\begin{figure*}
        \centering
        \begin{subfigure}[b]{0.475\textwidth}
            \centering
            \includegraphics[width=\textwidth]{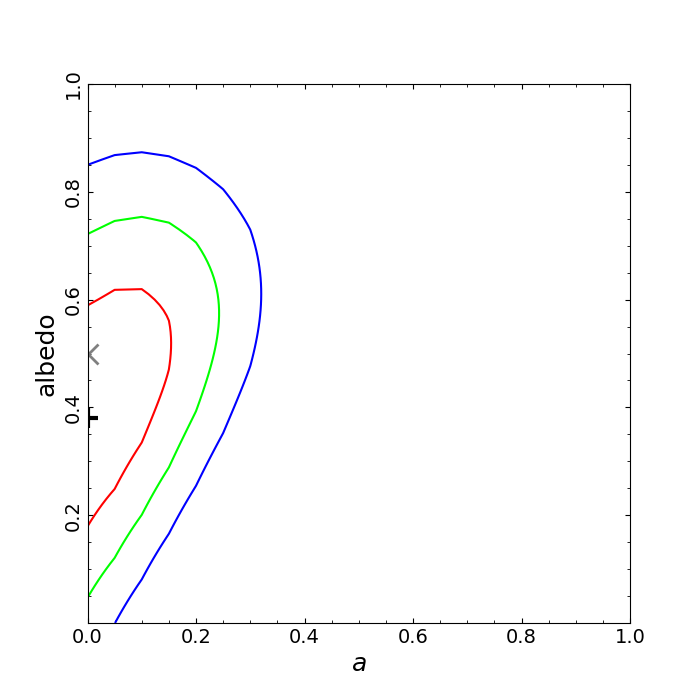}
            \caption[]%
            {{\small }}    
            \label{}
        \end{subfigure}
        \hfill
        \begin{subfigure}[b]{0.475\textwidth}  
            \centering 
            \includegraphics[width=\textwidth]{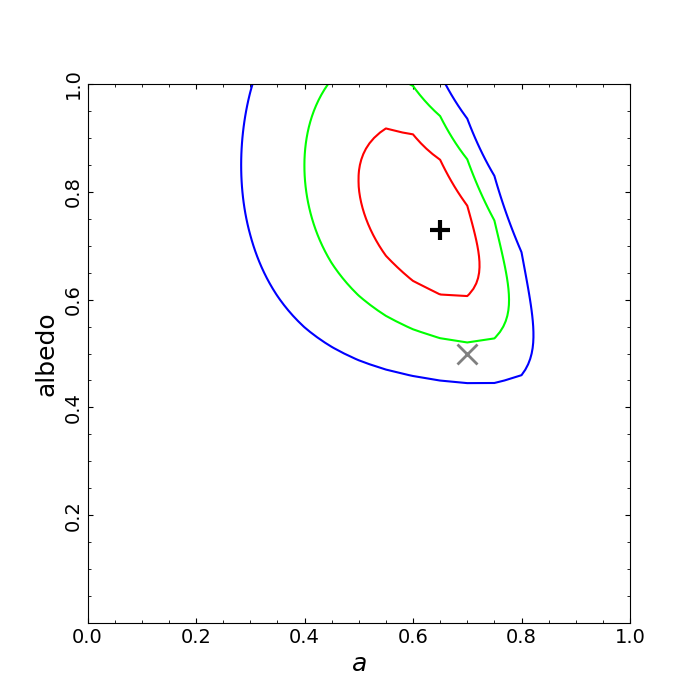}
            \caption[]%
            {{\small }}    
            \label{}
        \end{subfigure}
        \vskip\baselineskip
        \begin{subfigure}[b]{0.475\textwidth}   
            \centering 
            \includegraphics[width=\textwidth]{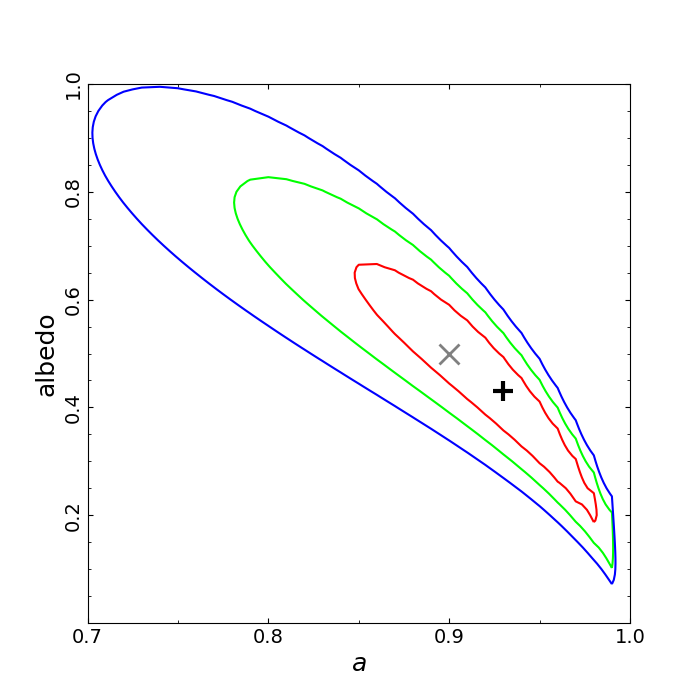}
            \caption[]%
            {{\small }}    
            \label{}
        \end{subfigure}
        \hfill
        \begin{subfigure}[b]{0.475\textwidth}   
            \centering 
            \includegraphics[width=\textwidth]{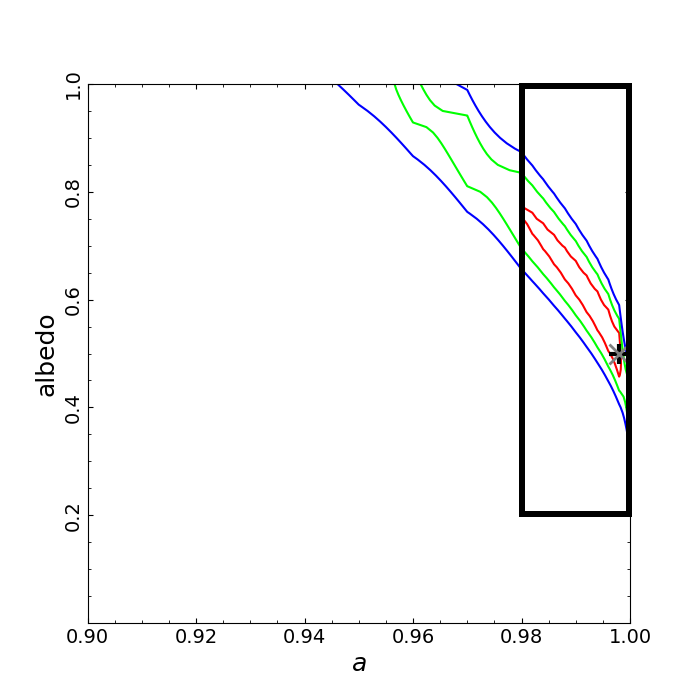}
            \caption[]%
            {{\small }}    
            \label{}
        \end{subfigure}
        \caption{Contour plots of spin vs albedo for spin values $a = 0$ (a), $0.7$ (b), $0.9$ (c) and $0.998$ (d) using simulated Stokes Q/E and U/E dependence on energy data. Plot in (d) is composed of two different grids divided by the black rectangular area, with the interval in spin $0.90$--$0.98$ having a step of 0.01 and the interval $0.98$--$1.00$ with the step of 0.002. In red 1$ \sigma $, green 2$ \sigma $ and blue 3$ \sigma $ contours. Similarly to the previous case, the grey cross sign marks the original value, which was 50 \% albedo in all the shown cases and spin a = 0 (a), a = 0.7 (b), a = 0.9 (c) and a = 0.998 (d). The black "plus" sign marks the best-fit value, which is as follows: a = 0, albedo = 38 \% (a), a = 0.65, albedo = 73 \% (b), a = 0.93, albedo = 43 \% (c), a = 0.998, albedo = 50 \% (d). Plots are calculated using a model with returning radiation (\textsc{kynbbrr}).}
        \label{fig:kynbbrr_contours}
\end{figure*}

We considered the presence of a less than 100\% albedo of the disk surface. In our simulations, we considered an energy-independent 50\% albedo. The Stokes parameters spectra, simulated for $a=0.998$ and for each of the three IXPE detector units and fit with theoretical model are shown in Figure \ref{fig:kynbbrr_data}. To address the capabilities in reproducing the observations, the contour plots of spin versus albedo were created. Since the code that we use for this study does not interpolate within spin values yet, the fitting procedure is as follows: we freeze all the parameters to their original (modeled) value, with the exception of albedo of the source. We perform a fit and thaw the spin parameter afterwards. We produce a contour plot of spin versus albedo. When a better fit for albedo is found, we change its value to the current best fit, freeze spin and perform fit followed by a contour spin-albedo calculation. We iterate this procedure until we find the fit with the lowest chi square statistics. Due to the interpolation issue, we used three different grids in spin for the calculation of the contour plots. They are as follows: $0$--$1$ interval with step 0.005 (subplots (a) and (b) of Figure \ref{fig:kynbbrr_contours}), $0.7$--$1$ with step of 0.01 (subplot (c) and the part $0.9$--$98$ of subplot (d)), and $0.98$--$1$ with the step of 0.002 ($0.98$--$1$ interval in the black rectangular area of the subplot (d)). We were able to successfully reconstruct the spin in all four of the studied cases (Figure \ref{fig:kynbbrr_contours}) - the gray cross (original value) and black "plus" signs (best-fit) overlap in spin (for $a = 0$ and $0.998$), or lay only a slight distance from one another (cases $a = 0.7$ and $0.9$). Similarly, albedo reconstruction was quite successful, although with rather large errors. Within 1$ \sigma $, the interval of satisfactorily fitting albedo values is approx. $ \Delta $ albedo = 0.45 for $a = 0$ and $0.9$ and approx. $ \Delta $ albedo = 0.3 for the cases $a = 0.7$ and $a = 0.998$.

\section{Discussion and conclusions} \label{sec:discussion}

We studied the robustness of constraining BH spin from the simulated Stokes parameters spectra. For a simulated exposure time of $500$ ks, we were able to successfully recover spin and inclination of the studied system for spin cases of $a = 0.7$, $0.9$ and $0.998$, for both the models with and without returning radiation. On the other hand, there is a strong degeneracy between the system inclination and the BH spin particularly for $a = 0$ in the model without the returning radiation. It will be therefore crucial to obtain inclination of the studied source independently, e.g. from spectroscopic observations.

Constraints on spin and inclination presented in this work were performed assuming distance, mass and accretion rate as known parameters. It is difficult to determine spin and inclination with either distance, mass or mass accretion rate as variable quantities, and this is the reason why we obtain a wider interval of possible values for the examined parameters and some spin cases might become indistinguishable.

In our initial studies we did not obtain a satisfactory fit for a spin - accretion rate case, suggesting $\dot{M}$ may not be sensitive on the polarization properties. However, we expect that the value of $\dot{M}$ will be constrained by the complementary information on the flux spectral shape that will be automatically provided by spectral analysis, which can be performed as well using the data collected by polarimetric observatories (like IXPE) or by coordinated observations with other facilities. This information, together with the polarimetric observations, will allow one to infer more precisely the fundamental X-ray Binary properties. X-ray polarimetry represents a highly desirable tool that will be beneficial when studying strong gravity effects of compact objects because it provides independent constraints on some of the parameters.

It is apparent from our results that considering the returning radiation effects adds more degeneracy to the problem, for example with the addition of the albedo parameter. In this case, to make things simpler, we treated the inclination as a known parameter. We obtained an acceptable fit for both $a = 0$ and $0.9$, although with large errors, as the corresponding contours cover almost half of the $0$--$1$ interval for the albedo. This implies that the detection of the albedo is strongly influenced by the spin of the black hole; this indicates that ignoring the albedo value will manifest itself in unconstrained spin. For $a = 0.7$ and $0.998$ we do not observe this degeneracy.

It is important to note that in this study we considered a constant albedo value across the energy spectrum. This can be fulfilled under the condition that the material in the accretion disk is fully ionized. In the recent work by \cite{Taverna2021}, this issue is generalized through studying the top layer of the accretion disk and computing its ionization profile, for which the polarization properties are obtained. Considering varying albedo profile thus allows for studying absorption, alongside scattering, as a process responsible for polarization. They find higher absorption linked to higher polarization degree. We plan to include all the above calculations in future version of the code.

Finally, it is worth to remark that eXTP will have on board a polarimeter with characteristics very similar to that of IXPE, but with an effective area 4-5 times larger. Our results can then be considered valid also for eXTP, once the exposure time has been rescaled accordingly.

\section*{Acknowledgements}

\noindent
RM and GM acknowledge financial support from the Italian Space Agency (grant 2017-12-H.0).
MB, MD and JS thank for the support from the GACR project 21-06825X.

\section*{Data Availability}

The data files used for the modeling and analysis presented in this paper were simulated using \textsc{kynbb}, \textsc{kynbbrr} and \textsc{IXPEobssim} software. \textsc{kynbb} code is publicly available at \url{https://projects.asu.cas.cz/stronggravity/kyn/}. \textsc{kynbbrr} is not publicly available, yet. For the simulation of the observations with IXPE, we used the \textsc{IXPEobssim} framework, which is available at \url{https://github.com/lucabaldini/ixpeobssim}.
 



\bibliographystyle{mnras}








\bsp	
\label{lastpage}
\end{document}